\documentclass[english]{elsarticle}
\usepackage[T1]{fontenc}
\usepackage[latin9]{inputenc}
\usepackage{prettyref}
\usepackage{amsmath}
\usepackage{amssymb}
\usepackage{graphicx}
\usepackage{esint}

\makeatletter
\usepackage[english]{babel}

\makeatother

\usepackage{babel}
\begin{document}

\title{Probabilistic simulation of mesoscopic ``Schr\"odinger cat'' states}

\author{B. Opanchuk}

\author{L. Rosales-Z\'arate}

\author{M. D. Reid}

\author{P. D. Drummond}

\address{Centre for Atom Optics and Ultrafast Spectroscopy, Swinburne University
of Technology, Melbourne 3122, Australia}

\ead{pdrummond@swin.edu.au, Tel: }
\begin{abstract}
We carry out probabilistic phase-space sampling of mesoscopic Schr\"odinger
cat quantum states, demonstrating multipartite Bell violations for
up to 60 qubits. We use states similar to those generated in photonic
and ion-trap experiments. These results show that mesoscopic quantum
superpositions are directly accessible to probabilistic sampling,
and we analyze the properties of sampling errors. We also demonstrate
dynamical simulation of super-decoherence in ion traps. Our computer
simulations can be either exponentially faster or slower than experiment,
depending on the correlations measured.
\end{abstract}
\maketitle
The calculation of quantum dynamics for large quantum systems is exponentially
hard if carried out through direct solutions of Schr\"odinger's equation.
In analyzing this problem, Feynman realized that probabilistic methods
could potentially solve this problem. Analyzing this further, he asked:
\emph{``Can quantum systems be probabilistically simulated by a classical
computer?''} \citep{Feynman1982}, and gave the answer \emph{``If
\ldots{} there's no hocus-pocus, the answer is certainly, No!''}.
Feynman continued, \emph{``This is called the hidden-variable problem:
it is impossible to represent the results of quantum mechanics with
a classical universal device''. }His argument apparently rested on
the assumption that Bell inequalities \citep{Bell1964} could not
be violated using probability distributions \citep{Hey1999,Stolze2008};
the remainder of his paper analyzed Clauser's two-photon Bell state
experiment \citep{Clauser1969} to illustrate the point.

Here we investigate Feynman's claim that probabilistic simulation
of quantum systems is impossible by providing, we believe, an important
explicit counterexample. We carry out a probabilitistic simulation
of the moments of an extreme Schr\"odinger cat quantum superposition
state: namely, the $N$-partite Greenberger-Horne-Zeilinger (GHZ)
state \citep{Greenberger1989}, that has been shown to violate multipartite
Bell inequalities for all $N$ \citep{Mermin1990}. For $N=2$, this
corresponds to the Bell state example given in Feynman's argument.
We investigate the scaling behavior of probabilistic sampling with
the number of particles $N$ through direct simulation of the multipartite
Bell violations. To achieve this, we employ the positive phase-space
distributions of quantum optics \citep{Husimi1940,Drummond1980,Hillery1984,Gardiner2004}
for our calculations. These have statistical moments which correspond
to those measured in Bell violations. 

Our study conveys the qualification necessary for Feynman's argument
to be strictly correct. Probabilistic simulation of quantum systems
by classical computer cannot be ruled out, based on the hidden variable
problem. Indeed, our results show that probabilistic quantum simulation
methods are promising. What is ruled out is only probabilistic simulation
based on the hidden variables of classical realism, where the trajectories
in phase space are constrained to correspond to the real values obtained
if the spins, positions and momenta of the particles were actually
measured. In our simulations, the variables associated with the simulation
are necessarily nonclassical: as in weak measurements \citep{Aharonov1988,Goggin2011},
their values extend outside the allowed eigenvalue range.

The methods we use are applied to the multi-qubit GHZ states with
up to $60$ spins (i.e. $N=60$), demonstrating macroscopic entanglement~\citep{Einstein1935}.
As well as being nonclassical, the quantum states we simulate have
Hilbert space dimension $10^{15}$ times larger than the 6 qubit universal
quantum computers~\citep{Lanyon2011} available currently. We show
that these methods can simulate the dynamics of super-decoherence
\citep{Monz2011}, which is relevant to a range of new, mesoscopic
quantum devices.

Our mesoscopic Schr\"odinger cat state simulations correspond to
states generated in recent ion-trap experiments with $N$ qubits or
ions \citep{Leibfried2005}. These states violate a genuine multipartite
Bell inequality, which means that it is impossible to confine the
Bell violation to just part of the system. We regard this as a worst-case
scenario for probabilistic simulations, as it not only demonstrates
a mesoscopic superposition, but gives the largest possible violation
of a Bell inequality. Such inequalities require the measurement of
all possible correlation functions at the highest order available.
We find two distinct scaling laws for the total computational difficulty,
as measured by the number of samples required to obtain a given sampling
error. 

For low order correlations, the number of samples required \emph{decreases
}with system-size. By contrast, to obtain all $N$-th order correlations,
an exponential \emph{increase} is found with a $2^{2N/3}$ power law,
due to increased sampling errors. However, as the distinct correlation
operators don't commute, an experimentalist would need $2^{N}$ measurements,
which is exponentially greater still. Surprisingly, therefore, a simulation
which calculates all observables simultaneously can be exponentially
faster than a experiment. This \emph{classical} parallelism more than
compensates for the sampling error in a probabilistic simulation. 

We first recall the definition of a Bell inequality. This is a constraint
on observable correlations of a physical system that obeys a local
hidden variable theory (LHV)~\citep{Bell1964,Clauser1969}. In the
multipartite case, the LHV theory must generate measurements by $N$
spatially separated observers, obtained from random samples of a parameter
$\lambda$. The measured values are functions of local detector settings
and the hidden parameter $\lambda$. The value observed by the $j$-th
observer with detector setting $a_{j}$ is $X_{j}(a_{j},\lambda)$.
All correlations are obtained from a probabilistic calculation of
the form:
\begin{equation}
\mathbf{C}(\mathbf{X})=\int_{\Lambda}\left[\prod_{j}X_{j}(\lambda,a_{j})\right]P(\lambda)d\lambda\label{eq:LHV}
\end{equation}

Here $X_{j}$ are experimental values, usually encoded as either $1$
or $-1$ in a binary experiment. These assumptions lead to inequalities
that any LHV correlations must satisfy. Genuine multipartite Bell
inequalities are the strongest Bell violations known, ruling out LHV
explanations for any subset of the observations. Quantum mechanics
is known to violate these inequalities, thus ruling out LHV theories.
But can one simulate these violations using probabilistic methods
\emph{equivalent} to quantum mechanics? 

In order to simulate Bell violations for multipartite spin or qubit
states, we use the SU(2)-Q distribution, which is well-suited to this
type of Hilbert space~\citep{Husimi1940,Arecchi1972}. Here one defines
\begin{equation}
Q\left(\mathbf{z}\right)=Tr\left[\hat{\rho}\hat{P}\left(\mathbf{z}\right)\right],
\end{equation}
where $\hat{P}\left(\mathbf{z}\right)$ is proportional to a coherent
state projection operator. This has the form of nonorthogonal, universal
POVM~\citep{DAriano2004}, with a corresponding physical measurement
strategy~\citep{Leonhardt1993}. The Q function calculation for a
spin expectation value is a sampled, probabilistic average over a
complex function $\sigma_{x}\left(z_{1}\right)$: 
\begin{equation}
\left\langle \hat{\sigma}_{x}^{1}\right\rangle =\left\langle \sigma_{x}\left(z_{1}\right)\right\rangle _{Q}=\frac{3}{2}\int d^{2N}\mathbf{z}\left(\frac{z_{1}+z_{1}^{*}}{1+\left|z_{1}\right|^{2}}\right)Q(\mathbf{z}).
\end{equation}
Full computational details and sampling methods will be given elsewhere.

To understand the ultimate scaling properties of such probabilistic
techniques, we have simulated the $N$-th order correlations that
violate multipartite Bell inequalities. These are found in quantum
states that display Bell violations with $N$ observers, not just
two. The most well-known examples are the multimode entangled Greenberger-Horne-Zeilinger
(GHZ) states~\citep{Greenberger1989}, generalized to $N$ spins
by Mermin\citep{Mermin1990} so that they are an example of an extreme
macroscopic superposition or ``Schr\"odinger Cat''. We considered
GHZ states which describe $N$ spin-$\frac{1}{2}$ particles or qubits:
\begin{equation}
\vert\Phi\rangle=\frac{1}{\sqrt{2}}\left(\vert\uparrow\ldots\uparrow\rangle+e^{i\phi}\vert\downarrow\ldots\downarrow\rangle\right).\label{eq:GHZ-state}
\end{equation}

Here $\left|\uparrow\right\rangle $ and $\left|\downarrow\right\rangle $
denote spin-up or spin down particles in the $z$-direction. As well
as being of deep significance in quantum physics, such mesoscopic
states have been generated in recent ion-trap experiments~\citep{Monz2011,Leibfried2005,Rowe2001}.
Quantum Bell inequality violations are obtained on measuring an operator
$\hat{A}$ which is defined as a linear combination of $2^{N}$ distinct
$N$-th order correlation functions:

\begin{equation}
\hat{A}=\prod_{j=1}^{N}\left(\hat{\sigma}_{x}^{j}+i\hat{\sigma}_{y}^{j}\right).\label{eq:GHZ-operator}
\end{equation}

For odd $N$ we follow Mermin~\citep{Mermin1990} and take $\phi=\pi/2$,
measuring $F=\Im\left\langle \hat{A}\right\rangle $, while for even
$N$ we follow Ardehali~\citep{Ardehali1992} and take $\phi=\pi$,
measuring $F=-\Re\left\langle \hat{A}\right\rangle $. Genuine multipartite
Bell violations are known to exist for these states, which imply that
the observed correlations cannot be explained by a nonlocality shared
among $N-1$ or fewer qubits. To test for genuine multipartite Bell
nonlocality, we adopt the hybrid local-nonlocal LHV model introduced
by Svetlichny~\citep{Svetlichny1987} and Collins et al.~\citep{Collins2002}. 

\begin{figure}
\begin{centering}
\includegraphics[width=0.75\columnwidth]{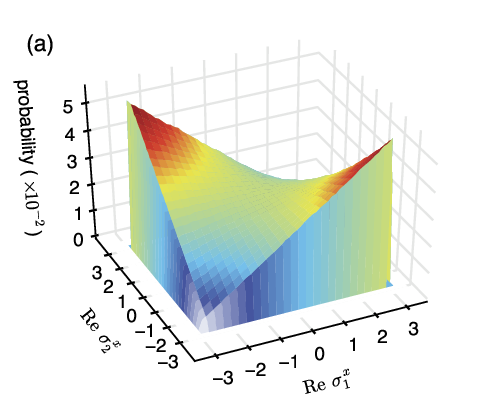}~\\
\includegraphics[width=0.75\columnwidth]{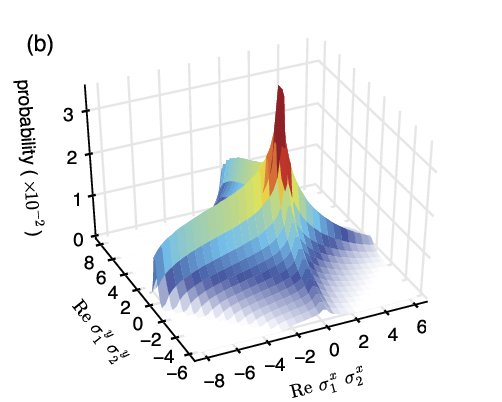}
\par\end{centering}

\caption{Correlations for the different parts of the quantity~\eqref{eq:F-2-particles},
in case of an SU(2) Q representation with two observers and $10^{8}$
samples.\label{fig:Spin-moments}}
\end{figure}

The difference between the SU(2) Q representation, compared to an
LHV theory, can be illustrated by considering the case of $N=2$,
where simplifying the Ardehali inequality will give rise to the corresponding
quantum mechanical expectation value:
\begin{equation}
F=-\langle\hat{\sigma}_{x}^{1}\hat{\sigma}_{x}^{2}\rangle+\langle\hat{\sigma}_{y}^{1}\hat{\sigma}_{y}^{2}\rangle=-\langle\sigma_{x}^{1}\sigma_{x}^{2}\rangle_{Q}+\langle\sigma_{y}^{1}\sigma_{y}^{2}\rangle_{Q}.\label{eq:F-2-particles}
\end{equation}
The correlation between real parts of the values of two factors for
one of the terms is plotted in~Fig.~\ref{fig:Spin-moments}(a),
and the correlation between the two terms of~\eqref{eq:F-2-particles}
is plotted in~Fig.~\ref{fig:Spin-moments}(b). Note that the SU(2)
representation does not limit the values of $\mathrm{Re}\,\sigma_{x}^{1}$
and $\mathrm{Re}\,\sigma_{x}^{2}$ to the range $[-1,1]$, as would
happen in an LHV theory; instead, they are bounded to $[-3,3]$. This
essential feature means that Bell's theorem does not restrict our
results, because the sampled values are not the same as their physical
eigenvalues. This property of having a different apparent range from
the operator eigenvalues, with the possibility of complex outcomes,
is also shared by the theory of weak measurements~\citep{Aharonov1988}. 

The high-dimensional multipartite state~\eqref{eq:GHZ-state} can
be readily sampled using probabilistic random number generators together
with the Q-function. 
\begin{figure}
\centering{}\includegraphics[width=0.75\columnwidth]{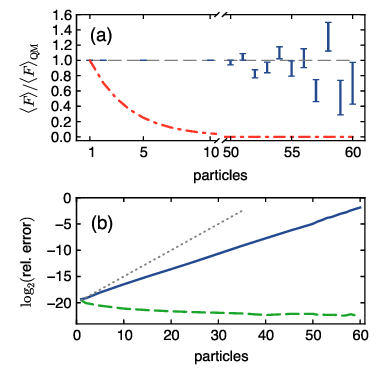}\caption{Violations for multi-particle GHZ states. (a) Simulated Mermin violation
using SU(2)-Q representation. The values of expectations and errors
are normalized by the quantum mechanical prediction for the corresponding
$N$. The horizontal grey dashed line gives the quantum prediction.
The error bars show the sampled result and estimated sampling errors
at each value of $N$. The red dash-dotted line is the LHV prediction,
which gives a Bell violation when above this line. Genuine multipartite
Bell violations occur for $F_{\mathrm{sample}}/F_{\mathrm{QM}}>1/\sqrt{2}$.
(b) Relative errors for $F$ (blue line) and first order correlation,
or total number of ``spin-ups'' (green dashed line) using SU(2)-Q
representation. The dotted reference line shows the point at which
the sampling errors would give scaling properties as slow as an experimental
measurement.\label{fig:GHZ-violations}}
\end{figure}
 These samples can then be used to calculate required expectation
values~(Fig.~\ref{fig:GHZ-violations}(a)). In the graph, the red
dashed line is the minimum correlation required to demonstrate a Bell
violation. Since the calculation is essentially a parallel one, we
employed graphical processor unit (GPU) technology to calculate many
samples in parallel, with a number of qubits ranging from $N=2$ to
$N=60$. This corresponds to measurement of a quintillion ($10^{18}$)
distinct sixtieth order correlation functions. Bell violations were
verified in all cases, while genuine multipartite violations of LHV
requiring all $N$ observers to participate, with $F_{sample}/F_{QM}>1/\sqrt{2}$,
were verified for $N<50$.

To understand the source of sampling errors, we investigated the scaling
of errors with system-size for\emph{ single} measurements of a low-order
spin correlation~(Fig.~\ref{fig:GHZ-violations}(b)), in addition
to the exponentially large numbers of measurements in the expectation
values $F$. As an example of low-order correlation we have chosen
the total number of ``spin-ups'' $N_{\uparrow}=\langle\sum_{j=1}^{N}\left(\hat{\sigma}_{z}^{j}+1\right)/2\rangle$.
Low-order correlations were easily calculated with decreasing\emph{
}sampling errors as $N$ increases. 

In contrast to this, high-order correlations showed an exponentially
increasing sampling error. The relative error in $F$ scales as $2^{N/3}$,
so that the time taken at constant error scales as $2^{2N/3}$. Hence,
probabilistic sampling scales more favorably than experiment, which
would take time proportional to $2^{N}$ due to the use of exponentially
many measurement settings. Single high-order correlations might be
faster, depending on experimental noise levels.

These results are unexpected and interesting. Sampling errors in low
order correlations are insensitive to scaling up to mesoscopic sizes.
These are the most commonly measured quantities, and there is no barrier
to sampling these, despite Bell violations and even mesoscopic superpositions.
However, correlations of the same order as the system size can be
exponentially hard to sample in the case of Schr\"odinger cat-like
states. Yet even in this case, the probabilistic strategy gives advantages.
It generates exponentially many non-commuting measurement results
in parallel, which can result in scaling that is exponentially faster
than with direct measurements. We emphasize that these scaling results
are specific to the GHZ case, and will change as the quantum state
is changed.

\begin{figure}
\centering{}\includegraphics[width=0.75\columnwidth]{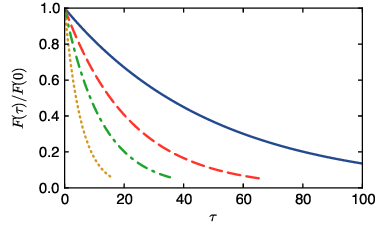}\caption{Decay of the sampled quantity $F$ after the application of the simple
model of super-decoherence~\prettyref{eq:H-decoherence}, for 2 (solid
blue line), 3 (red dashed line), 4 (green dash-dotted line) and 6
(yellow dotted line) particles, with decoherence rate $\epsilon=0.1$.
The horizontal axis is the dimensionless time, $\tau=t/\Delta t$.\label{fig:Decoherence}}
\end{figure}

In this paper, we have focused on the specific issue of whether one
can probabilistically sample the observables of a quantum state that
violates a Bell inequality. The larger problem of carrying out quantum
dynamical simulations was not treated, but encouraging results are
known. As an example, we consider the question of dynamical noise
and decoherence in ion traps, which is an important issue in the observation
of mesoscopic quantum effects~\citep{Brune1996}. The experimentally
observed magnetic field noise found in ion-trap experiments can be
easily added to our calculations using a simple model. Following Monz
et al.~\citep{Monz2011}, we assume a delta-correlated noise such
that $\left\langle \Delta B(t)\Delta B(t')\right\rangle =\Delta B_{0}^{2}\delta(t-t')$,
with an interaction Hamiltonian of the form:

\begin{equation}
\hat{H}=\frac{\mu\Delta B(t)}{2}\sum_{j=1}^{N}\hat{\sigma}_{z}^{j}.\label{eq:H-decoherence}
\end{equation}
We find that this can also be readily simulated dynamically. This
was achieved by multiplying an independent noise term $\exp\left(i\epsilon N\zeta_{j}\right)$
by the value corresponding to the operator $A$ in each of the samples
after every time step $\Delta t$. Here $\epsilon=\mu\Delta B_{0}\sqrt{\Delta t}/\hbar$
defines the speed of the decoherence, and $\zeta_{j}$ is a Gaussian
random number such that $\left\langle \zeta_{j}\zeta_{j'}\right\rangle =\delta_{jj'}$.
This is shown in~Fig.~\ref{fig:Decoherence}, which demonstrates
the experimentally observed quadratic super-decoherence as $N$ increases.

Other examples of dynamical quantum correlations have been treated
using phase-space methods, including quantum soliton dynamics~\citep{Drummond1993a},
interacting quantum fields~\citep{Deuar2007} equivalent to $\sim10^{6}$
qubits and the Dicke superfluorescence model~\citep{Altland2012}.
These simulated correlations are in general agreement with experimental
observations~\citep{Jaskula2010}. This shows the wide range of potential
applicability of these techniques. However, the optimal dynamical
methods and overall questions of efficiency are not yet known. We
emphasize that these examples have limitations. Our results show that
probabilistic methods are not ruled out, rather than giving an optimum
recipe for dynamical quantum simulations. 

In summary, we show that low order quantum correlations are the simplest
to obtain with probabilistic sampling. Higher order correlations in
GHZ states \emph{can} also be simulated, but with greater difficulty.
While these require exponentially many samples to reduce errors to
acceptable levels, they do not require exponentially large memory
resources. In his paper, Feynman proposed the development of universal
quantum computers to solve this problem, and these could potentially
carry out simulations~\citep{Lloyd1996}. However, this proposed
hardware is currently limited in size to $6$ qubits or less~\citep{Lanyon2011},
and has proved difficult to scale to large size. It remains fundamentally
useful to obtain theoretical predictions with software that doesn't
require new technology, both for practical and scientific reasons.
After all, if we wish to \emph{test} quantum theory on mesoscopic
scales, we cannot assume that our hardware strictly obeys quantum
mechanics on large scales.

Importantly, we demonstrate that probabilistic digital algorithms
on existing hardware can simulate mesoscopic quantum superpositions
much larger than any current experiment. Such technologies are proposed
for quantum secret-sharing~\citep{Hillery1999} and high-precision
atom interferometers~\citep{He2011}, amongst others. Digital simulations
could therefore have a direct application both to fundamental physics,
and to the design and implementation of these devices. Our demonstration
that mesoscopic Bell violations \emph{can }be treated probabilistically
will lead to further developments. The fact that our simulations can
be exponentially \emph{faster} than direct measurements is a surprising
consequence of the classical parallelism inherent in this computational
strategy. While Feynman might regard our approach as `hocus-pocus',
it is certainly probabilistic.

\subsection*{Acknowledgements}

L. E. C. R. Z. acknowledges financial support from CONACYT, Mexico.
P. D. D. and M. D. R. acknowledge discussions with P. Deuar, and the
Australian Research Council for funding via a Discovery grant.

\bibliographystyle{elsarticle-num}
\bibliography{BellQLsimsLetter}

\end{document}